\documentclass[prl,aps,twocolumn,superscriptaddress,preprintnumbers,amsmath,amssymb]{revtex4}
\usepackage[dvips]{graphicx}% Include figure files
\usepackage{dcolumn}% Align table columns on decimal point
\usepackage{bm}% bold math

\newcommand{\be}{\begin{equation}}
\newcommand{\la}{\langle}
\newcommand{\ra}{\rangle}
\newcommand{\ee}{\end{equation}}
\newcommand{\bea}{\begin{eqnarray}}
\newcommand{\eea}{\end{eqnarray}}
\newcommand{\bd}{\begin{displaymath}}
\newcommand{\ed}{\end{displaymath}}
\renewcommand{\a}{\alpha}

\newcommand*{\origr}{}
\let\origr\r
\renewcommand{\r}{\rho}
\newcommand{\s}{\sigma}

\renewcommand{\d}{\delta}

\newcommand{\vf}{\varphi}

\renewcommand{\o}{\omega}

\newcommand{\G}{\Gamma}
\newcommand{\lf}{\left}
\newcommand{\rg}{\right}

\begin{document}

\title{Decoherence in adiabatic quantum evolution -- application to Cooper pair pumping}
\author{J. P. Pekola}
\affiliation{Low Temperature Laboratory, Aalto University,
P.O.~Box 13500, FI-00076 AALTO, Finland}
\author{V. Brosco}
\affiliation{Dipartimento di Fisica, Universita' di Roma "La
Sapienza", P.le A. Moro, 2 00185 Roma, Italy} 
\affiliation{\hbox{Institut f\"ur Theoretische Festk\"orperphysik, Karlsruhe Institute of Technology, 76128 Karlsruhe, Germany}
}
\affiliation{ISC-CNR,  Via dei Taurini, 19 00185 Roma, Italy } 
\author{M. M\"ott\"onen}
\affiliation{Low Temperature Laboratory, Aalto University,
P.O.~Box 13500, FI-00076 AALTO, Finland}
\affiliation{Department of Applied Physics/COMP, Aalto University, P.O.~Box 15100, FI-00076 AALTO, Finland}
\affiliation{Australian Research Council Centre of Excellence for Quantum Computer
    Technology, School of Electrical Engineering \& Telecommunications,
    University of New South Wales, Sydney, New South Wales 2052, Australia}
\author{P. Solinas}
\affiliation{Department of Applied Physics/COMP, Aalto University,
P.O.~Box 15100, FI-00076 AALTO, Finland}
\author{A. Shnirman}
\affiliation{\hbox{Institut f\"ur Theorie der Kondensierten Materie,
Karlsruhe Institute of Technology, 76128 Karlsruhe, Germany}}
\affiliation{\hbox{DFG Center for Functional Nanostructures (CFN),
Karlsruhe Institute of Technology, 76128 Karlsruhe, Germany}}

\begin{abstract}
One of the challenges of adiabatic control theory is the proper inclusion
of the effects of dissipation. Here, we study the adiabatic dynamics of an
open two-level quantum system
deriving a generalized master equation to consistently account for the
combined action of the driving and dissipation. We demonstrate that in
the zero temperature limit the ground state dynamics is not affected by environment. As an example, we apply our theory to Cooper pair pumping which demonstrates the robustness of ground
state adiabatic evolution.
\end{abstract}
\maketitle %{\sl Introduction.}
Accurate control of quantum systems
has been one of the greatest challenges in physics for the last
decades. Adiabatic temporal evolution~\cite{berry84} has attracted a
lot of attention \cite{falci00,farhi01,bacon09,rezakhani09} in this respect since it provides robustness
against timing errors and typically utilizes evolution in the ground
state of the system. Such evolution has been argued to be
robust against relaxation and environmental noise \cite{childs02,florio06,fubini07}.

The combined effect of adiabatic evolution and dissipation
were considered by many authors using various techniques and with different aims and assumptions,
see, e.g., Refs.~\cite{childs02,sarandy05,thunstrom05,wubs09,carollo03,whitney05}.
We derive in this Letter a unique master equation that treats the combined effect of noise and adiabatic
driving consistently and, thus, provides a pioneering tool for studying the effects of decoherence in quantum control protocols employing adiabaticity ~\cite{farhi01,bacon09}.
We find that adiabatic evolution should not be treated in the secular
approximation~\cite{cohentannoudji}.
Furthermore, the master equation incorporates new terms
ensuring relaxation into the correct time-dependent ground state. When these issues are properly addressed, the expectation values of physical observables in the adiabatically steered ground state are not influenced
by zero-temperature dissipation.
We apply our theory to adiabatic charge transport in superconducting circuits in the presence of noise.
In spite of its long history \cite{thouless, geerligs91, averin98, jp99},
this problem
%has recently attracted a revived interest due
%to its fundamental relation to geometric \cite{aunola03, governale05, mottonen06, brosco08,mottonen08} and
has recently attracted revived theoretical \cite{governale05, mottonen06, brosco08,hekking03} and experimental \cite{vartiainen07,mottonen08} interest due to its fundamental relation to geometric \cite{aunola03} and
topological \cite{leone} phases and to
its potential applications in metrology \cite{vartiainen07,niskanen03}.

We consider an open quantum system subject to external time-dependent control fields. The total Hamiltonian of the system and its environment, $H(t)$, is the sum of three terms,
$H(t)=H_S(t)+H_E+ V$, where $H_S(t)$ denotes the time-dependent system Hamiltonian, $H_E$ is the bath Hamiltonian and $V$ is the system-bath coupling. Assuming that the driving does not directly affect the coupling term between the system and the environment, we can write $V=X \otimes Y$, where $X$ is a bath operator and $Y$ is a system operator.
In the case of weak system-noise coupling and slow driving, a convenient basis to describe the dynamics of the
system is the instantaneous energy eigenstate basis, also called \emph{adiabatic basis}, defined by
$H_S(t)\lf|\psi_n(t)\rg>=E_{n}(t)\lf|\psi_n(t)\rg>$.
The states $|\psi_n(t)\ra$ are assumed to be normalized and non-degenerate. We denote by $D(t)$ the transformation from a given fixed basis to the adiabatic one.
%; $D^{\dag}(t)D(t)=D(t)D^{\dag}(t)=I$.
The evolution of the transformed density matrix is governed by the effective Hamiltonian
\begin{equation}\label{ham1}
\tilde H^{(1)}(t)=\tilde H_S(t)+ \hbar w(t) + \tilde V(t)+ H_E
\end{equation}
where $\tilde H_S(t)= D^\dag(t)H_S(t)D(t)$, $\tilde{V}(t)=D^\dag(t) V D(t)=X\otimes \tilde Y(t)$, and $w = -iD^\dagger\dot{D}$.
%arises from the classical control parameters via the rotation $\hat{U}$ that diagonalizes the Hamiltonian,

We note that there are a few possible strategies of treating the dissipation. The usual one
is to disregard $w$ in the calculation of the dissipative rates~\cite{childs02}. Then, zero-temperature environment
tends to relax the system to the ground state of $H_S(t)$, while the rotation $w$ tries
to excite the system. The resulting state is different from both the adiabatic ground state (ground state of $H_S$)
and from the ground state of $\tilde H_S+\hbar w$. The second
strategy is to first perform a series of transformations to the super-adiabatic bases~\cite{Berry_Superadiabatic,whitney05} and, then treat the dissipation. The first step
would be to diagonalize $\tilde H_S+\hbar w$ with a unitary transformation $D_1$ and get a much smaller
non-adiabatic correction $w_1 = -iD_1^\dagger\dot{D_1}$. Here, the dissipation (treated in Markov approximation)
takes us to the ground state of $\tilde H_S+\hbar w$.
Although not exact, the second strategy allows one to treat the combined
effect of noise and driving consistently.
Here we adopt this strategy to calculate the lowest order
correction to the adiabatic dissipative dynamics of a two-level system. As
we will show, up to higher order corrections, this treatment correctly
accounts for the relaxation to the ground state of the superadiabatic
hamiltonian $\tilde H_S+\hbar w$. By using standard methods explained, e.g.,
in Ref. \cite{cohentannoudji}, we arrive at the following master equation for the reduced system density
matrix
$\tilde{\rho}_I(t)$ in the interaction picture (for the derivation see Appendix):%\ref{Appendix}:
\begin{widetext}
\begin{eqnarray} \label{me1}
\frac{d\tilde{\rho}_I(t)}{dt}=&& i
[\tilde{\rho}_I(t),w_I(t)]-\frac{1}{\hbar^2}{\rm Tr}_E \lf\{\int_0^t dt'
\Big[\big[\tilde{\rho}_I(t)\otimes\r_E,\tilde{V}_I(t')\big],\tilde{V}_I(t)\Big]\rg\}\nonumber \\ && +\frac{i}{\hbar^2}{\rm Tr}_E\lf\{\int_0^t dt'\int_{0}^{t'} dt''
\Big[\big[\tilde{\rho}_I(t)\otimes\r_E,[w_I(t'),\tilde{V}_I(t'')]\big],\tilde{V}_I(t)\Big]\rg\},
\end{eqnarray}
\end{widetext}
where ${\rm Tr}_E$ indicates trace over the
environmental degrees of freedom and ${\rho}_E$ is the stationary density operator of the environment. To obtain Eq. (2) we have to take consistently into account corrections up to the order $wVV$, resulting in a nonstandard commutator expression. The interaction picture operators are defined as
$\tilde O_I(t)=e^{i H_E t/\hbar} U^\dag_S(t,0)\tilde O (t)U_S(t,0)e^{-i H_E t/\hbar}$,
where $U_S(t,0)= e^{-i \int_0^{t}\tilde H_S(\tau)d\tau/\hbar}$ is the system time-evolution operator.
%$\tilde{V}_I(\tau)=(\tilde{H}_E)\tau/\hbar}\tilde{V}(\tau)e^{-i(\tilde{H}_S+\tilde{H}_E)\tau/\hbar}$,
%Above, we have employed the interaction picture such that $\hat{V}_I(\tau)=e^{i(\hat{H}_S+\hat{H}_E)\tau/\hbar}\hat{V}(\tau)e^{-i(\hat{H}_S+\hat{H}_E)\tau/\hbar}$, $\hat{w}_I(\tau)=e^{i\hat{H}_S\tau/\hbar}\hat{w}(\tau)e^{-i\hat{H}_S\tau/\hbar}$, and $\hat{\rho}_I(\tau)=e^{-i\hat{H}_S\tau/\hbar}\textrm{Tr}_E\left\{\hat{\rho}(\tau)\right\}e^{i\hat{H}_S\tau/\hbar}$.
%the operators $\hat O$, i.e., $\hat V$ and $\hat w$, take the form $\hat{O}_I(\tau)=e^{i\hat{H}_S\tau/\hbar}\hat{O}(\tau)e^{-i\hat{H}_S\tau/\hbar}$ and the density operator the form $\hat{\rho}_I(\tau)=e^{-i\hat{H}_S\tau/\hbar}\hat{\rho}(\tau)e^{i\hat{H}_S\tau/\hbar}$ in the interaction picture at time $\tau$.
In Eq. (\ref{me1}), the first contribution on the r.h.s. is of order $\alpha=\hbar/(\Delta T_p)$
where $\Delta$ is the minimum gap in the spectrum of $H_S$ and $T_p$ is the period
on which the Hamiltonian is varied \cite{alpha}.
The second term is as in the standard Bloch-Redfield theory. The third
one is a cross-term of the drive and dissipation ensuring relaxation to the
proper ground state \cite{whitney05}.

We now focus on the case of a general two-state system, with the instantaneous
eigenstates $|g\rangle$ (ground state) and $|e\rangle$ (excited
state).
In this case, returning to the Schr\"odinger
picture, we can recast Eq. (\ref{me1}) into
\begin{widetext}
\begin{eqnarray} \label{dc1}
\dot{\rho}_{gg}= &&-2\Im\mbox{m}(w_{ge}^*\rho_{ge})-(\Gamma_{ge}+\Gamma_{eg})\rho_{gg}+\Gamma_{eg}+{\tilde\Gamma}_0\Re\mbox{e}(\rho_{ge})\nonumber \\&&
+\frac{\Re\mbox{e} (w_{ge})}{\omega_{0}}[(2\tilde\Gamma_+ -\tilde\Gamma_0)(1-\rho_{gg})-(2\tilde\Gamma_- -\tilde\Gamma_0)\rho_{gg}]
+2\frac{\Re\mbox{e}(w_{ge})\Re\mbox{e}(\rho_{ge})}{\omega_{0}}(\Gamma_{ge}+\Gamma_{eg}-\Gamma_0),
\end{eqnarray}
and
\begin{eqnarray} \label{dc2}
\dot{\rho}_{ge}=&& iw_{ge}(2\rho_{gg}-1)+i(w_{ee}-w_{gg})\rho_{ge}+i\omega_{0}\rho_{ge}-
i(\Gamma_{ge}+\Gamma_{eg})\Im\mbox{m}(\rho_{ge})-\Gamma_\varphi
\rho_{ge}+(\tilde\Gamma_+
+\tilde\Gamma_-)\rho_{gg}-\tilde\Gamma_+\nonumber \\ &&
+\big[\frac{w_{ge}}{\omega_{0}}(2\Gamma_--\Gamma_\varphi)-i\frac{\Im\mbox{m}(w_{ge})}{\omega_{0}}(\Gamma_{eg}-\Gamma_{ge})\big]\rho_{gg}
-\big[\frac{w_{ge}}{\omega_{0}}(2\Gamma_+ -\Gamma_\varphi)+i\frac{\Im\mbox{m}(w_{ge})}{\omega_{0}}(\Gamma_{eg}-\Gamma_{ge})\big](1-\rho_{gg})\nonumber \\
&&+2\big[ \frac{w_{ge}^*}{\omega_{0}} \Re\mbox{e}(\rho_{ge})+2i\frac{\Re\mbox{e}(w_{ge})}{\omega_{0}}\Im\mbox{m}(\rho_{ge})\big](\tilde\Gamma_0-\tilde\Gamma_+-\tilde\Gamma_-).
\end{eqnarray}
\end{widetext}
%The over-dot refers to time derivative.
By $O_{kl}$ we denote the matrix elements $\la m|O|n\ra$ of a general operator $O$, with $m,n=e,g$, except $w_{mn}=-i\langle m|\dot n\rangle$. We have defined the rates
$\Gamma_{ge}=\frac{Y_{ge}^2}{\hbar^2}S(-\omega_{0})$
(excitation),
$\Gamma_{eg}=\frac{Y_{ge}^2}{\hbar^2}S(+\omega_{0})$
(relaxation), $\Gamma_{\varphi}=2\frac{Y_{gg}^2}{\hbar^2}S(0)$
(dephasing), and the less common transition terms
$\tilde{\Gamma}_\pm=\frac{Y_{gg}Y_{ge}}{\hbar^2}S(\pm\omega_{0})$,
$\tilde{\Gamma}_0=2\frac{Y_{gg}Y_{ge}}{\hbar^2}S(0)$,
$\Gamma_\pm=\frac{Y_{gg}^2}{\hbar^2}S(\pm\omega_{0})$, and
$\Gamma_0=2\frac{Y_{ge}^2}{\hbar^2}S(0)$. Here,
%$\hat{V}_I(\tau)=e^{i(\hat{H}_S+\hat{H}_E)\tau/\hbar}\hat{V}(\tau)e^{-i(\hat{H}_S+\hat{H}_E)\tau/\hbar}$
the matrix elements of $Y$ obey
$Y_{gg}(t)=-Y_{ee}(t)$ and $Y_{eg}(t)=Y_{ge}(t)$~\cite{sigmaz}. The energy separation between the two states is $\hbar\omega_0$, which varies along the pumping trajectory. The power spectrum of the noise is
defined through $S(\omega)=\int_{-\infty}^\infty\la X_{I}(\tau)X_I(0)\ra e^{i\o\tau}d\tau$.
%We disregard the imaginary parts
%of the rates, e.g., the Lamb shifts. These renormalization effects provided mainly by the high frequency environment should be included self-consistently into %the Hamiltonian of the system.

%$\hat{V}_I(\tau)=e^{i(\hat{H}_S+\hat{H}_E)\tau/\hbar}\hat{V}(\tau)e^{-i(\hat{H}_S+\hat{H}_E)\tau/\hbar}$. The. %We assume that both $\hat V$ and $\hat w$ are hermitian and traceless.
%Note that all this applies to a general two-level system, provided the traces of $\hat V$ and $\hat w$ vanish.
%In deriving Eqs. (\ref{dc1}) and \eqref{dc2} from Eq. (\ref{me1}),
Throughout, we have used Markov approximation, i.e., we neglect the variation of $\tilde \r_I(t)$ between $t$ and $t+\tau_c$, assuming that the correlation time of the bath, $\tau_c$, is much shorter than the typical relaxation time of the system, $1/\G$. Furthermore, we made the approximation of \emph{adiabatic rates} (AR), i.e., in the calculation of the rates we neglect the slow variation of $\omega_0$, $Y$, and $w$, assuming the bath correlation time to be much shorter than the driving period $\tau_c \ll T_p$.
On the other hand, Eqs. (\ref{dc1}) and \eqref{dc2} include all
the non-secular terms traditionally neglected~\cite{cohentannoudji}. They introduce cross-dependence between $\rho_{gg}$ and $\rho_{ge}$ in the dissipative terms, and in our problem, omitting them would lead to unphysical results, such as violation of charge conservation.

We are interested in the quasi-stationary limit that the system reaches when the evolution is adiabatic and
it is initially in the ground state.
We thus look for the solutions of $\dot{\rho}_{gg}=0$ and $\dot{\rho}_{ge}=0$ for $\alpha \ll 1$.
Since $w_{mn}=O(\alpha)$, in the absence of dissipation, we find that
$\r_{gg} \simeq 1+O(\a^2)$ and $\r_{ge}\simeq -w_{ge}/\o_0 + O(\a^2)$
are the desired solutions.
In the zero-temperature limit, $S(-\omega_{0})=0$, to the first order in $\a$, Eqs. \eqref{dc1} and \eqref{dc2} yield, again,
$\rho_{gg} = 1+O(\a^2)$ and the following equation for the off-diagonal element up to order $\a$:
%\begin{equation} \label{final2}
$i\omega_{0}\Omega_{ge}-\Gamma_\varphi \Omega_{ge}-i\Gamma_{eg}\Im\mbox{m}(\Omega_{ge}) = 0$,
%\end{equation}
%
with $\Omega_{ge}\equiv \rho_{ge}+w_{ge}/\omega_{0}$.
The solution of this equation is exactly the same as for the closed system; $\rho_{ge}=-w_{ge}/\omega_{0}$.
Therefore, the ground state evolution is not influenced by coupling to a zero-temperature Markovian environment in the adiabatic limit. Note, that including the imaginary part of the rates, e.g., the
Lamb shift, does not change this result.

The vanishing of the effects of dissipation is consistent with the following simple argument.
In the zero temperature limit, and to first order in $\a$, the effect of dissipation is to bring the system to the instantaneous ground state of the effective Hamiltonian, $\tilde H_1=\tilde H_S+\hbar w$, which means that in the eigenbasis of $\tilde H_1$ spanned by the eigenvectors, $|\tilde \psi^{(1)}_n\ra$, the density matrix has the form $\tilde \r^{(1)}_{mn}=\la\tilde \psi^{(1)}_m|\r\,|\tilde \psi^{(1)}_n\ra=\d_{mg}\d_{ng}+O(\a^2)$ independent of the dissipative rates.
Thus, within our approximations, the ground state evolution is robust against zero-temperature environmental noise
and the expectation value of any operator in the quasi-stationary evolution does not depend on the specific properties of the environment. If, instead, we neglect the non-secular terms, we obtain the same solution for $\rho_{gg}$ but the
evolution of $\rho_{ge}$ is influenced by the noise as $\rho_{ge}=-w_{ge}/(\omega_0+i\Gamma/2)$, where $\Gamma$ represents a combination of the dissipative rates.
This leads to different expectation values of physical observables that depend on $\rho_{ge}$, and
to the loss of robustness of the ground state dynamics. Therefore, in general, the non-secular terms cannot be neglected:
they give a leading order contribution in $\Gamma \alpha/\Delta$ to the
dynamics.

To test our theory on a concrete example, we discuss a superconducting Cooper pair pump.
It consists of an array of Josephson junctions coupled to two superconducting leads, being subject to time-dependent external fields. As discussed by various authors (see, e.g., Ref. \cite{hekking03}), the transferred charge is the sum of a dynamic and a geometric contribution, $Q=Q^D+Q^G$.
The first one corresponds to the average supercurrent and the
second one to pumping.
Assuming that only two levels are involved, the two contributions to the charge transferred through junction $i$ in a pumping cycle can be written as
\bea \label{currentd}
Q^D_i&=&\!\!\int_0^{T_p}\!\! (\rho_{gg}I_{i,gg}+\rho_{ee}I_{i,ee})dt,\\
Q^G_i&=&\!\!\int_0^{T_p} \!\!2\Re \mbox{e}(\rho_{ge}I_{i,eg}) dt \label{currentg},
\eea
where $\hat{I}_i$ is the current operator through junction $i$.
Here we focus on the pumped charge, i.e., $ Q^G_i \equiv\int_0^{T_p} \!I^{\rm G}_i dt$ \cite{footnote2}.
%%
%\be Q^{\rm G}_i\equiv \int_0^{T_p} \!\!\!I^{\rm G}_i dt= 2\!\!\int_0^{T_p} \!\!\!\!\Re \mbox{e}(\rho_{ge}I_{i,eg})dt.\ee
%%
%
\begin{figure}
\begin{center}
\includegraphics[width=7.5cm]{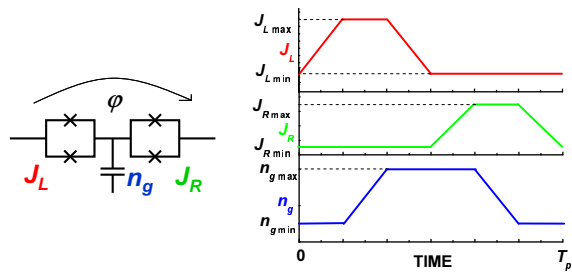}
\end{center}
\caption{An example of a quantum pump, the Cooper pair sluice, is shown on the left. A pumping
cycle is sketched on the right. The time-dependent classical control parameters are the magnetic fluxes tuning the Josephson tunnel couplings $J_L$ and $J_R$ and the gate voltage controlling the offset charge $n_g$ of the island. They vary in time with period $T_p$, whereas the phase difference across the device, $\varphi$, is stationary.}\label{scheme}
\end{figure}
By substituting $\r_{ge}= -w_{ge}/\o_0$ in Eq. \eqref{currentg} we arrive at the well-known formula for the adiabatically pumped current in a closed system \cite{jp99},
%\begin{equation} \label{pumpchi0}
$I^{\rm G}_i = -\frac{2}{\omega_{0}}\Re\mbox{e}(w_{ge}I_{i,eg})$.
%\end{equation}
As discussed above, this is also the limit of the adiabatic evolution in the presence of environmental noise.

\begin{figure}
\begin{center}
\includegraphics[width=7cm]{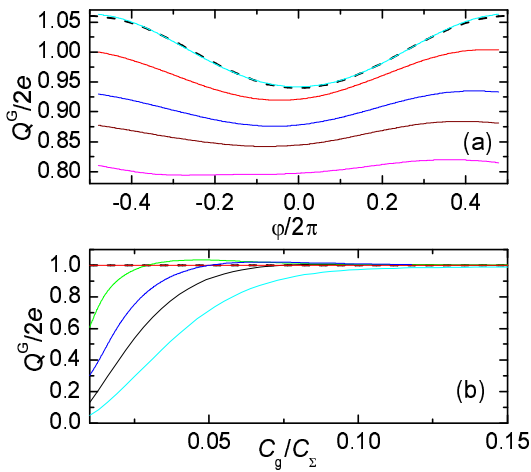}
\end{center}
\caption{Pumped charge of the sluice under gate charge noise. (a) Phase dependence of the pumped charge $Q^{\rm G}$ in a fully symmetric pumping cycle with respect to $J_L$ and $J_R$ of Fig. \ref{scheme} as a function of the phase bias $\varphi$. The dashed line shows the analytic result of Eq.~\eqref{analytic} for adiabatic pumping. The solid lines are from the numerical calculations based on Eqs.~\eqref{dc1} and \eqref{dc2} for $f\equiv T_p^{-1}=100$ MHz, with $C_g/C_\Sigma = 0.015,0.0175,0.02,0.025,$ and $0.3$ from bottom to top. (b) Coupling dependence of the pumped charge at $\varphi=\pi/2$. The dashed line shows the analytic result as in (a). The solid lines are for $f = 10,100,150,200$ and $300$ MHz from top to bottom. The other parameters are $J_{\rm max}/E_{\rm C}=0.1$, $J_{\rm min}/J_{\rm max}=0.03$, $n_{g\,{\rm max}}=0.8$, $n_{g\, {\rm min}}=0.2$, $E_{\rm C}/k_{\rm B}=1$ K ($E_{\rm C}/2\pi\hbar= 21$ GHz), $R=300$ k$\Omega$, environment temperature $T=0$, $S(\omega_
 0)=2\hbar\omega_{0}R$, $S(-\omega_0)=0$, and $S(0) = 2k_{\rm B}T_0R$ with $T_0=0.1$~K.}
\label{numerics}
\end{figure}

%{\sl Cooper pair sluice.}
In particular, we consider the Cooper pair
sluice \cite{niskanen03} of Fig.~\ref{scheme}.
It consists of a single superconducting island, coupled to two superconducting leads via two SQUIDs, i.e., Josephson junctions whose critical currents can be tuned by magnetic fluxes.
The electrostatic potential on the island can be controlled by a gate voltage, $V_g$, and
there is a constant superconducting phase difference, $\vf=\vf_L-\vf_R$ between the two leads.
In the absence of noise, the Hamiltonian of the
sluice can be written as
\begin{equation} \label{HS}
H_S =E_C(n-n_g)^2 -J_L \cos (\varphi_L - \theta) -
J_R \cos ( \theta-\varphi_R ).
\end{equation}
Here $\theta$ and $n$ are the operators for the superconducting phase of the island and
the number of excess Cooper pairs on it. The Josephson couplings to left and right lead are denoted as $J_L$ and $J_R$, $n_g=C_gV_g/2e$ is the normalized gate charge, and $E_{C}= 2e^2/C_{\Sigma}$ is the charging energy of the sluice; $C_g$ is the gate capacitance and
$C_{\Sigma}$ the total capacitance of the island.
The current operators of the left and right junctions read $I_L= \frac{2 e}{\hbar} J_L \sin (\varphi_L - \theta)$ and $I_R= \frac{2 e}{\hbar} J_R \sin ( \theta-\varphi_R )$, respectively.
For $E_{\rm C}\gg {\rm max}\{J_L,J_R\}$, and $n_g\simeq 1/2$ only two charge states, $|1\ra$ and $|0\ra$, i.e., one or no extra Cooper pairs on the island, are relevant.
Dissipation is then mostly due to gate voltage fluctuations.
Other noise sources, not considered here, are determined by fluctuations of the fluxes in the SQUIDs or in $\varphi$~\cite{hekking03}. In the two-level approximation, the coupling between sluice and charge-noise has the form
%\begin{equation} \label{Vg}
$V = - g \s_z\otimes \delta V_g(t)$,
%\end{equation}
where $g=eC_g/C_\Sigma$ is the coupling constant, $\s_z=|0\rangle\langle 0|-|1\rangle \langle
1|$ and $\delta V_g(t)$ is the gate voltage
fluctuation.
%In this basis the instantaneous eigenstates can be expressed as
%\bea
%|g\ra&=&e^{i\g}\cos \z |0\ra+\sin\z |1\ra \nn\\
%|e\ra&=&e^{i\g}\sin \z |0\ra-\cos\z |1\ra.
%\eea
%In the previous equation, the angle $\z$ and the phase $\g$ are given by:
% $2\z=\arctan(|J_Z|/\ve)$, $\g=\arg (J_Z)$
%and we set $J_Z=J_Le^{-i\vf/2}+J_Re^{i\vf/2}$ and $\ve=E_c(n_g-1)$.\\
%\begin{equation} \label{vi1}
%\end{equation}
%Let us start reviewing briefly the results for the non-dissipative sluice \cite{mottonen06}.
In the absence of dissipation, for the cycle of Fig. \ref{scheme}, with $J_{i} \in[J_{\rm min},J_{\rm max}]$, $n_g \in[n_{g\,{\rm min}},n_{g\,{\rm max}}]$ and for $J_{\rm max}\ll E_{\rm C}$, one obtains the pumped charge in the adiabatic limit according to Eq. \eqref{currentg} as
\begin{equation} \label{analytic}
Q^{\rm G}_i = 2e(1-2\frac{J_{\rm min}}{J_{\rm max}}\cos\varphi)
\end{equation}
for both junctions \cite{niskanen03}. Thus the transported charge depends on $\varphi$, the
average being one Cooper pair per cycle.
%This result does not depend
%appreciably on the extremes $n_{g\,{\rm min}},\,n_{g\,{\rm max}}$
%as long as $n_g$ crosses twice the position $n_g=1/2$, where the states $|0\ra$ and $|1\ra$ are degenerate.
In the presence of dissipation, Eqs.
\eqref{dc1} and \eqref{dc2} were integrated numerically to obtain the temporal
evolution of the density matrix along a pumping trajectory of
Fig.~\ref{scheme}. Figure \ref{numerics} shows that, upon increasing the system-environment coupling at finite frequencies $f\equiv T_p^{-1}$,
the pumped charge approaches the analytic result of Eq.~\eqref{analytic} for adiabatic pumping at all values of $\varphi$,
see Fig.~\ref{numerics}(a). Figure
\ref{numerics}(b) shows the coupling dependence of the
pumped charge at various frequencies for $\varphi=\pi/2$. On lowering
the frequency, all the data collapse towards the horizontal dashed
line which is again the result of Eq.~\eqref{analytic}. For $f=10$
MHz, the numerical and analytic results are indistinguishable on this
scale. Thus coupling to zero-temperature Markovian environment
seems to be useful for adiabatic ground state pumping.
We note, however, that Eqs. (\ref{dc1}--\ref{dc2}) are strictly valid only for adiabatic evolution and
weak coupling.
%the numerical integration of these equations
%thus provides only a hint on the dynamics beyond this regime.
%Furthermore, the Lamb shift neglected here but investigated in~\cite{whitney05}
%might induce some coupling dependence
%of the pumped charge.

In conclusion,
we derived a master equation for an adiabatically driven two-level system
including the combined effect of drive and relaxation. We found it important to account for the time-dependence of the Hamiltonian of the system in determining the dissipative rates and to include the non-secular terms.
As an example, we analyzed adiabatic Cooper pair pumping in the ground state and demonstrated
that the pumped charge is not influenced
by zero-temperature environment.
Numerical solution of
the master equation suggests that dissipation can resume adiabatic pumping at finite frequencies.

We thank R. Fazio for many very useful discussions. We have received funding from the European Community's
Seventh Framework Programme under Grant Agreement No. 238345 (GEOMDISS). MM acknowledges Academy of Finland and Emil Aaltonen Foundation for financial support.

\appendix
\label{Appendix}
\section{Derivation of the master equation}

%\section{Derivation of the expression for the adiabatically pumped charge}
The master equation~(\ref{me1}) is obtained with a development taking into account all the terms up to the order $w V V$.
As in standard derivation of the master equation \cite{cohentannoudji2}, we assume that the density matrix of the environment is stationary and that the average of $V$ over the environment degrees of freedom vanishes.

Denoting the total density matrix of the system and  environment as $\tilde{\rho}^{\rm tot}(t)$ and employing the transformation to the adiabatic basis as $\tilde{\rho}_I^{\rm tot}=D^\dagger \tilde{\rho}^{\rm tot} D$, the von Neumann equation in the interaction picture reads
\begin{equation} \label{eq1}
\dot{\tilde{\rho}}_I^{\rm tot}(t)=\frac{i}{\hbar}[\tilde{\rho}_I^{\rm tot}(t), \hbar w_I(t)+\tilde{V}_I(t)].
\end{equation}
Notice that, in the weak coupling and in the adiabatic limit, $w_I(t)$ and $\tilde{V}_I(t)$ are perturbative contributions of different order.
Tracing over the degree of freedom of the environment, Eq.~(\ref{eq1}) becomes
\begin{equation}
\dot{\tilde{\rho}}_I(t)=i [\tilde{\rho}_I(t), w_I(t)]+ \frac{i}{\hbar} {\rm Tr}_E \big{\{}[\tilde{\rho}_I^{\rm tot}(t),\tilde{V}_I(t)]\big{\}},
\label{eq:eq2}
\end{equation}
where $\tilde{\rho}_I^{\rm tot}= {\rm Tr}_E \big{\{} \tilde{\rho}_I^{\rm tot} \big{\}}$.

Together with Eq.~(\ref{eq1}), we employ the identity
\begin{equation}
\tilde{\rho}_I^{\rm tot}(\tau)=\tilde{\rho}_I^{\rm tot}(\tau_1)+\int_{\tau_1}^\tau d\tau' \dot{\tilde{\rho}}_I^{\rm tot}(\tau').
\label{eq:formal_rho}
\end{equation}
Using iteratively Eqs.~(\ref{eq1}) and (\ref{eq:formal_rho}), we can obtain a perturbation expansion of Eq.~(\ref{eq:eq2}).

Substituting Eq.~(\ref{eq:formal_rho}) in the last term in Eq.~(\ref{eq:eq2}) we have
\begin{eqnarray}
&&\dot{\tilde{\rho}}_I(t)=i[\tilde{\rho}_I(t), w_I(t)]+\nonumber \\
&&\frac{i}{\hbar} {\rm Tr}_E \big{\{}[\tilde{\rho}_I^{\rm tot}(0),\tilde{V}_I(t)]
+\int_0^t dt'[\dot{\tilde{\rho}}_I^{\rm tot}(t'),\tilde{V}_I(t)]\big{\}}.
\end{eqnarray}

Since the average of $V$ over the environment degrees of freedom vanishes,  ${\rm Tr}_E \big{\{}[\tilde{\rho}_I^{\rm tot}(0),\tilde{V}_I(t)] \big{\}}=0$ and using Eq.~(\ref{eq1}), we obtain
\begin{widetext}
\begin{eqnarray}
\dot{\tilde{\rho}}_I(t)&=&i [\tilde{\rho}_I(t), w_I(t)]-\frac{1}{\hbar^2} {\rm Tr}_E
\big{\{}\int_0^t dt'\big[ [\tilde{\rho}_I^{\rm tot} (t'),\hbar w_I(t')],\tilde{V}_I(t)\big]\big{\}}
-\frac{1}{\hbar^2} {\rm Tr}_E\big{\{}\int_0^t dt'\big[ [\tilde{\rho}_I^{\rm tot} (t'),\tilde{V}_I(t')],\tilde{V}_I(t)\big]\big{\}}.
%=\nonumber \\
%&=&\frac{i}{\hbar} [\tilde{\rho}_I(t), \hbar w(t)] -\frac{1}{\hbar^2} {\rm Tr}_E\big{\{} \int_0^t dt'\big[[\int_0^{t'}dt'' \dot{\tilde{\rho}}_I^{\rm tot} (t''), \hbar w(t')],\tilde{V}_I(t)\big]+\int_0^t dt'\big[[\tilde{\rho}_I^{\rm tot}(t),\tilde{V}_I(t')],\tilde{V}_I(t)\big]\nonumber \\
%&&-\int_0^t dt'\big[[\int_{t'}^t dt'' \dot{\tilde{\rho}}_I^{\rm tot}(t''),\tilde{V}_I(t')],\tilde{V}_I(t)\big]\big{\}},
\end{eqnarray}
\end{widetext}
The term $\tilde{\rho}_I^{\rm tot} (t')$ can be transformed using Eq. (\ref{eq:formal_rho}); in particular, we substitute
$\tilde{\rho}_I^{\rm tot} (t')=\tilde{\rho}_I^{\rm tot} (0)+\int_{0}^{t'} dt^{\prime \prime} \dot{\tilde{\rho}}_I^{\rm tot} (t^{\prime \prime} )$  and $\tilde{\rho}_I^{\rm tot} (t')=\tilde{\rho}_I^{\rm tot} (t)-\int_{t'}^{t} dt^{\prime \prime}
\dot{\tilde{\rho}}_I^{\rm tot} (t^{\prime \prime} )$ in the second and in the third term on the right, respectively.
Consistently, we have that ${\rm Tr}_E \{ \big[[\tilde{\rho}_I^{\rm tot}(0),w_I(t')],\tilde{V}_I(t)\big]\} =0$ and hence
\begin{widetext}
\begin{eqnarray}
\dot{\tilde{\rho}}_I(t)&=&i  [\tilde{\rho}_I(t), w_I(t)]
-\frac{1}{\hbar^2} {\rm Tr}_E\big{\{} \int_0^t dt'\big[[\tilde{\rho}_I^{\rm tot}(t),\tilde{V}_I(t')],\tilde{V}_I(t)\big]\big{\}}
-\frac{1}{\hbar^2} {\rm Tr}_E\big{\{} \int_0^t dt' \int_0^{t'}dt'' \big[[ \dot{\tilde{\rho}}_I^{\rm tot} (t''), \hbar w_I(t')],\tilde{V}_I(t)\big]\big{\}}\nonumber \\
&&+\frac{1}{\hbar^2} {\rm Tr}_E\big{\{}  \int_0^t dt' \int_{t'}^{t} dt'' \big[[ \dot{\tilde{\rho}}_I^{\rm tot}(t''),\tilde{V}_I(t')],\tilde{V}_I(t)\big]\big{\}},
\end{eqnarray}
\end{widetext}

Using Eq. (\ref{eq1}) we eliminate $\dot{\tilde{\rho}}_I^{\rm tot}(t'')$ from the above equation and, keeping the terms up to order $w V V$, we obtain
\begin{widetext}
\begin{eqnarray}
\dot{\tilde{\rho}}_I(t)&=&i[\tilde{\rho}_I(t), w_I(t)]
-\frac{1}{\hbar^2} {\rm Tr}_E\big{\{} \int_0^t dt'\big[[\tilde{\rho}_I^{\rm tot}(t),\tilde{V}_I(t')],\tilde{V}_I(t)\big]\big{\}}
-\frac{i}{\hbar^2} {\rm Tr}_E\big{\{} \int_0^t dt' \int_0^{t'}dt'' \Big[ \big[ [\tilde{\rho}_I^{\rm tot} (t''),\tilde{V}_I(t'')], w_I(t')\big],\tilde{V}_I(t)\Big]\big{\}}\nonumber \\
&&+\frac{i}{\hbar^2} {\rm Tr}_E\big{\{}  \int_0^t dt' \int_{t'}^t dt'' \Big[ \big[ [ \tilde{\rho}_I^{\rm tot}(t''),  w_I(t'') ],\tilde{V}_I(t') \big],\tilde{V}_I(t)\Big]\big{\}},
\label{eq:last_eq}
\end{eqnarray}
\end{widetext}

The third and fourth terms on the right are both of order $w V V$, namely, the highest order in our expansion.
The last step in our derivation is to use Eq.~(\ref{eq:formal_rho}) to substitute $\tilde{\rho}_I^{\rm tot}(t'')$ with $\tilde{\rho}_I^{\rm tot} (t)+\int_{t}^{t''} dt'''
\dot{\tilde{\rho}}_I^{\rm tot} (t''')$; however, since the terms with derivative of $\tilde{\rho}_I^{\rm tot}$ give contributions either of order $w$ or $V$, they can be neglected.
Thus, we can effectively substitute $\tilde{\rho}_I^{\rm tot}(t'')$ with  $\tilde{\rho}_I^{\rm tot}(t) $ in Eq.~(\ref{eq:last_eq}) without introducing further approximations.
The master equation~(\ref{me1}) is obtained by rearranging the integration limits and the commutators of the last two terms of the resulting equation.

\section{Alternative Derivation of the master equation (diagrammatic approach)}

\begin{figure*}
\begin{center}
    \includegraphics[width=5cm]{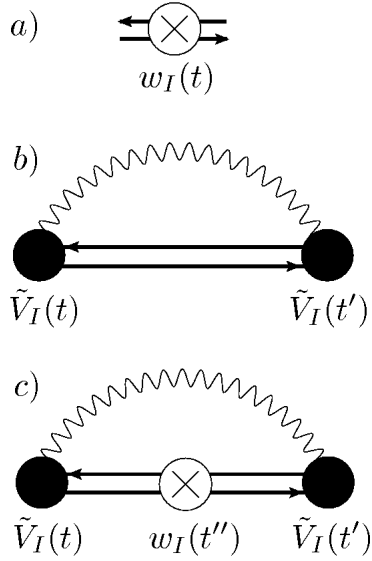}
    \end{center}
    \caption{Three diagrams contributing to the self-energy. The wavy line corresponds to the
    bath correlators $\langle X(t)X(t')\rangle$ and $\langle X(t')X(t)\rangle$. The horizontal lines
    are the pieces of the Keldysh contour. The vertices $\tilde V_I$ and $w_I$ can be put onto
    the upper (with prefactor $-i$) or the lower (with prefactor $i$) Keldysh line. This renders the
    commutators in Eq.~(\ref{eq:nonMarkovME}). In explicit diagrammatic
    notations of Ref.~\cite{SchoellerSchoen}
    the diagram a) corresponds to two diagrams, the diagram b) to four, and
    diagram c) to eight.}
    \label{Fig:diagr}
\end{figure*}

A standard iterative derivation of the master equation \cite{cohentannoudji2} can be cast
into a diagrammatic form \cite{SchoellerSchoen} which allows for a systematic accounting of the
higher order terms. We have two different perturbations, $w_I(t)$ and $\tilde{V}_I(t)$, and
we perform an expansion taking into account all the terms up to the order $w V V$.
According to \cite{SchoellerSchoen} the master equation reads
\begin{equation}
\dot{\tilde{\rho}}_I(t) = \int_0^{t} dt' \Sigma_I(t,t') \tilde{\rho}_I(t')\ ,
\end{equation}
where $\Sigma_I$ is the self-energy given by the sum of all irreducible diagrams.
The irreducible diagrams relevant for the order $w V V$ are shown in Fig.~\ref{Fig:diagr}.
This produces
the following non-Markovian master equation:
\begin{widetext}
\begin{eqnarray}
\dot{\tilde{\rho}}_I(t)&=&-i[w_I(t), \tilde{\rho}_I(t)]-\frac{1}{\hbar^2} {\rm Tr}_E \int_0^t dt'\big[\tilde{V}_I(t),[\tilde{V}_I(t'),\tilde{\rho}_I(t')\otimes \rho_E]\big]
\nonumber\\
&+&\frac{i}{\hbar^2}{\rm Tr}_E \int_0^t dt' \int_{t'}^{t}dt'' \Big[\tilde{V}_I(t), \big[w_I(t''), [\tilde{V}_I(t'),\tilde{\rho}_I(t')\otimes \rho_E]\big]\Big]\ .
\label{eq:nonMarkovME}
\end{eqnarray}
\end{widetext}
The three terms in the RHS of (\ref{eq:nonMarkovME}) correspond to the diagrams a), b), and c) of Fig.~(\ref{Fig:diagr}) respectively.

A naive Markovian approximation would be to substitute $\tilde{\rho}_I(t')$ by $\tilde{\rho}_I(t)$ in both the second and the third terms of the RHS of (\ref{eq:nonMarkovME}). We, however, note, that a Markovian
approximation consistent with keeping the third term of the RHS of (\ref{eq:nonMarkovME}) requires a substitution of the following expression into the second term of the RHS of (\ref{eq:nonMarkovME}):
\begin{equation}
{\tilde{\rho}}_I(t') \approx {\tilde{\rho}}_I(t) + i \int_{t'}^{t} dt'' [w_I(t''), \tilde{\rho}_I(t'')]\ .
\end{equation}
That is, in making the Markovian approximation in the second term of the RHS of (\ref{eq:nonMarkovME}),
we have to take into account the correction provided by the first term of the RHS of (\ref{eq:nonMarkovME}).
We obtain
\begin{widetext}
\begin{eqnarray}
\dot{\tilde{\rho}}_I(t)&=&-i[w_I(t), \tilde{\rho}_I(t)]-\frac{1}{\hbar^2} {\rm Tr}_E \int_0^t dt'\big[\tilde{V}_I(t),[\tilde{V}_I(t'),\tilde{\rho}_I(t)\otimes \rho_E]\big]
\nonumber\\
&+&\frac{i}{\hbar^2}{\rm Tr}_E \int_0^t dt' \int_{t'}^{t}dt'' \Big[\tilde{V}_I(t), \big[w_I(t''), [\tilde{V}_I(t'),\tilde{\rho}_I(t')\otimes \rho_E]\big]\Big]\nonumber\\
&-&\frac{i}{\hbar^2}{\rm Tr}_E \int_0^t dt' \int_{t'}^{t}dt'' \Big[\tilde{V}_I(t), \big[\tilde{V}_I(t'), [w_I(t''),\tilde{\rho}_I(t'')\otimes \rho_E]\big]\Big]\ .
\label{eq:nonMarkovMEcorrected}
\end{eqnarray}
\end{widetext}
Finally, upon making the Markovian approximation in the third and the fourth
terms of the RHS of (\ref{eq:nonMarkovMEcorrected}), i.e., by substituting
$\tilde{\rho}_I(t')$ by $\tilde{\rho}_I(t)$ and $\tilde{\rho}_I(t'')$ by $\tilde{\rho}_I(t)$ and rearranging
the commutators we obtain the following master equation
\begin{widetext}
\begin{eqnarray}
\dot{\tilde{\rho}}_I(t)&=&-i[w_I(t), \tilde{\rho}_I(t)]-\frac{1}{\hbar^2} {\rm Tr}_E \int_0^t dt'\big[\tilde{V}_I(t),[\tilde{V}_I(t'),\tilde{\rho}_I(t)\otimes \rho_E]\big]
\nonumber\\
&+&\frac{i}{\hbar^2}{\rm Tr}_E \int_0^t dt' \int_{t'}^{t}dt'' \Big[\tilde{V}_I(t), \big[[w_I(t''),\tilde{V}_I(t')],\tilde{\rho}_I(t)\otimes \rho_E\big]\Big]\ .
\label{eq:nonMarkovMEcorrected}
\end{eqnarray}
\end{widetext}


\begin{thebibliography}{99}
\bibitem{alpha} Momentarily, adiabaticity is governed by $|w|/\omega_0$, with $|w|$ denoting the norm of $w$, and $\hbar\omega_0$ the level separation.
\bibitem{sigmaz} We assume $Y_{ge}$ to be real, which applies for the gate voltage fluctuations analyzed here.
\bibitem{footnote2}The dynamical contributions can be experimentally separated from the geometric ones as discussed in Refs. \cite{mottonen06, mottonen08}.
\end{thebibliography}
\end{document}